\documentclass[11pt,twoside]{article}
\usepackage{asp2010}

\resetcounters

\begin{document}

\title{A grid of MARCS model atmospheres for S stars}
\author{Sophie~Van Eck$^1$, Pieter~Neyskens$^1$, Bertrand~Plez$^2$,
  Alain~Jorissen$^1$, Bengt Edvardsson$^3$, Kjell Eriksson$^3$, Bengt
Gustafsson$^3$, Uffe-Gr\aa e J\o rgensen$^4$ and \AA ke Nordlund$^4$}
\affil{$^1$Institut d'Astronomie et d'Astrophysique, Universit\'e Libre de
  Bruxelles, Boulevard du Triomphe, B-1050 Brussels, Belgium}
\affil{$^2$GRAAL, Universit\'e de Montpellier II, CNRS-UMR 5024, Place Eug\`ene Bataillon, F-34095 Montpellier cedex 5, France}
\affil{$^3$Department of Astronomy and Space Physics, Uppsala Astronomical Observatory, box 515, S-751~20 Uppsala, Sweden}
\affil{$^4$ Niels Bohr Institute for Astronomy, Physics and Geophysics, Copenhagen University, Blegdamsveg 17, 
Copenhagen, DK-2100 Denmark}

\begin{abstract}
S-type stars are late-type giants whose atmosphere is enriched in carbon
and s-process elements because of either extrinsic pollution by a binary
companion or intrinsic nucleosynthesis and dredge-up on the
thermally-pulsing AGB. A large grid of S-star
model atmospheres has been computed covering the range  $2700 \le T_{\rm
  eff}(\rm K) \le 4000$ with $0.5 \le {\rm C/O} \le 0.99$.
ZrO and TiO band strength indices as well as VJHKL
photometry are needed to disentangle T$_{\rm
  eff}$, C/O and [s/Fe]. 
A ``best-model finding tool'' was developed using a set of well-chosen indices and checked against
photometry as well as low- and high-resolution spectroscopy.  
It is found that applying M-star model atmospheres (i.e., with a solar C/O
ratio) to S stars can lead to errors on T$_{\rm  eff}$ up to 400~K. 
We constrain the
parameter space occupied by S stars of the vast sample of Henize stars in terms of T$_{\rm  eff}$, [C/O] and [s/Fe].

\end{abstract}

\section{Introduction}
The S class was originally defined by \cite{Merrill-22} to designate a group of
curious red stars which did not fit well into either class M (TiO stars) or
classes R and N (carbon stars). \cite{Keenan-54} clarified the situation by
accepting  as S stars only those  exhibiting ZrO bands. 
The numerous attempts to link phenomenological spectral classification
criteria to physical parameters (T$_{\rm eff}$, gravity, C/O, [s/Fe],
[Fe/H]) \citep{Keenan-54,Keenan-McNeil-76,Ake-79,Keenan-Boeshaar-80} only
lead to imprecise results, because low-resolution diagnostics are
strongly entangled in terms of T$_{\rm eff}$, C/O and [s/Fe] variations.
The only in-depth discussion of the thermal structure dates
back to the pioneering paper of \cite{Piccirillo-80}. He already insisted on
the strong influence of the C/O ratio on the atmospheric structure and
spectra of S stars, in addition to effects due to the s-process elements
overabundance. His investigation was however mostly limited to qualitative
statements, due to obvious technical limitations. Most subsequent analysis
of S stars relied on models designed for M-type stars, not allowing for C/O or [s/Fe] ratio changes.
In the present paper we present a new grid of
model atmospheres, superseeding the one presented in \cite{Plez-03}, covering most of the parameter space of S-type stars, and
attempt to provide a calibration of photometric indices in terms of T$_{\rm eff}$, C/O and [s/Fe].

\section{Model atmospheres and spectra}
\begin{figure}[!ht]
\label{Fig:struct-model}
\plotfiddle{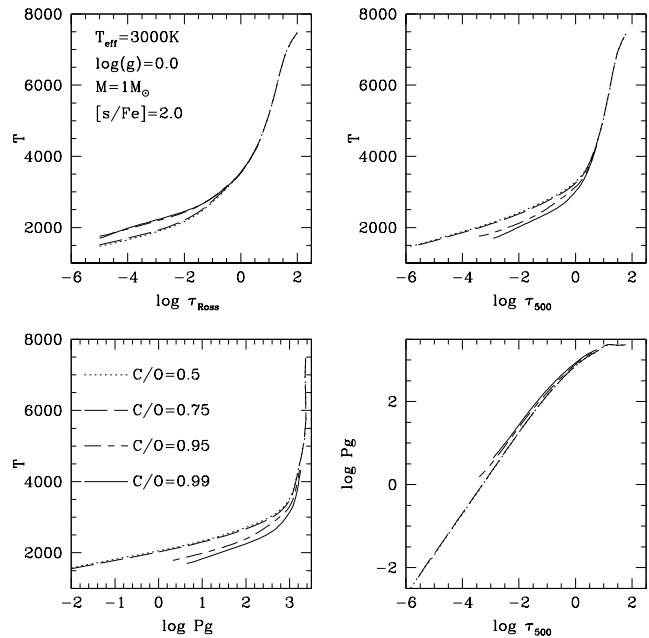}{7.0cm}{0}{45}{45}{-140}{-80}
\caption{Thermal and  pressure structure of models with T$_{\rm eff}$ =
  3000 K and [s/Fe]=+2 dex}
\end{figure}

Since models for S star atmospheres are virtually non-existent, a grid
of MARCS model atmospheres (see \cite{Gustafsson-08} for details on the
models computation) for S stars has been calculated: 
 $2700 \le {\rm
  T}_{\rm eff} $ (K) $ \le 4000$ (step of 100K); 
 C/O = 0.5, 0.750,
0.899, 0.925, 0.951, 0.971, 0.991; 
 $[\rm{s}/\rm{Fe}]$ = 0., +1., +2. dex ; 
 $[\rm{Fe}/\rm{H}]$ = -0.5 and  0. dex ;  
 log(g) = 0,1,2,3,4,5.

All models were computed for M = 1 M$_\odot$ and with $[\alpha /\rm{Fe}] =
-0.4 \times [\rm{Fe}/\rm{H}]$. Opacities as complete and accurate as
possible were included, including polyatomic  molecules and a specific ZrO linelist (described in Plez et al., in
  preparation). Models were computed through opacity sampling with more than
  10$^5$ wavelength points, local thermodynamic equilibrium, mixing-length
  theory of convection and
  spherical symmetry for log(g) $\le 2$.

\begin{figure}[!ht]
\label{Fig:model-CO}
\plotfiddle{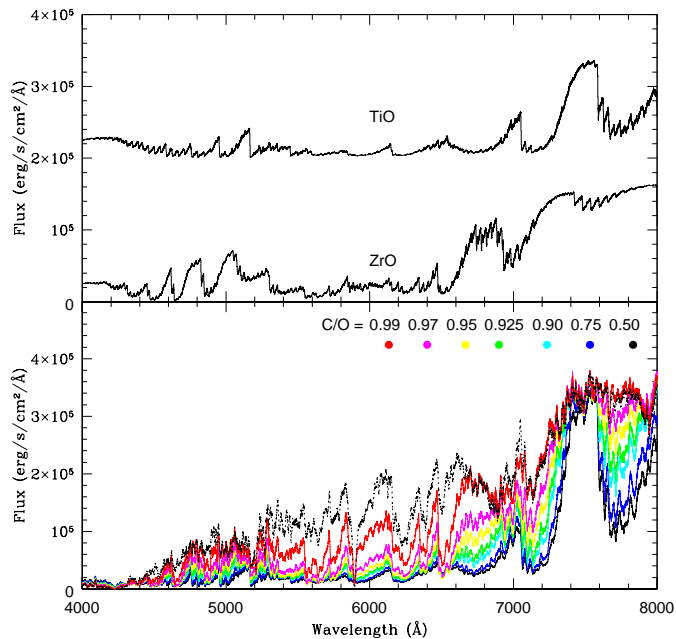}{7.5cm}{0}{45}{45}{-140}{-75}
\caption{Top: absorption spectra of TiO and ZrO at 3000~K. Bottom: Thick
  lines: spectra
  with increasing C/O ratios (from bottom to top) for
  T$_{\rm eff}$=3000~K and [s/Fe]=+1 dex. Thin dashed line: spectrum with
  C/O=0.99 but no s-process enhancement ([s/Fe]=0 dex)}
\end{figure}

A total of 3522 converged model atmospheres was obtained. The model
structure for T$_{\rm eff}$=3000 K and [s/Fe]=+2 dex models is shown in Fig.~1, where the major
influence of C/O on the thermal structure is readily apparent (whereas the
[s/Fe] ratio has less importance). The P$_{\rm gas}$ - $\tau_{500}$ relation
(fixed mostly by log g) stays basically unchanged, whereas the T -
$\tau_{\rm Ross}$ relation (governed by the energy balance requirement) reaches
higher temperatures at the surface for higher C/O. When C/O increases,
P$_{\rm gas}$ at a given T increases. The latter effects are due to a large
decrease of the partial pressures of H$_2$O and TiO, two major opacity contributors.
Fig.~\ref{Fig:model-CO} illustrates how the depth of TiO and ZrO bands decreases with
increasing C/O, for different models of T$_{\rm eff}$=3000~K, and the
influence on the spectra of the level of s-process enhancement.

\section{Confronting the models with observed color
and spectral band indices}
\begin{figure}[!ht]
\label{Fig:TiOZrO}
\plotfiddle{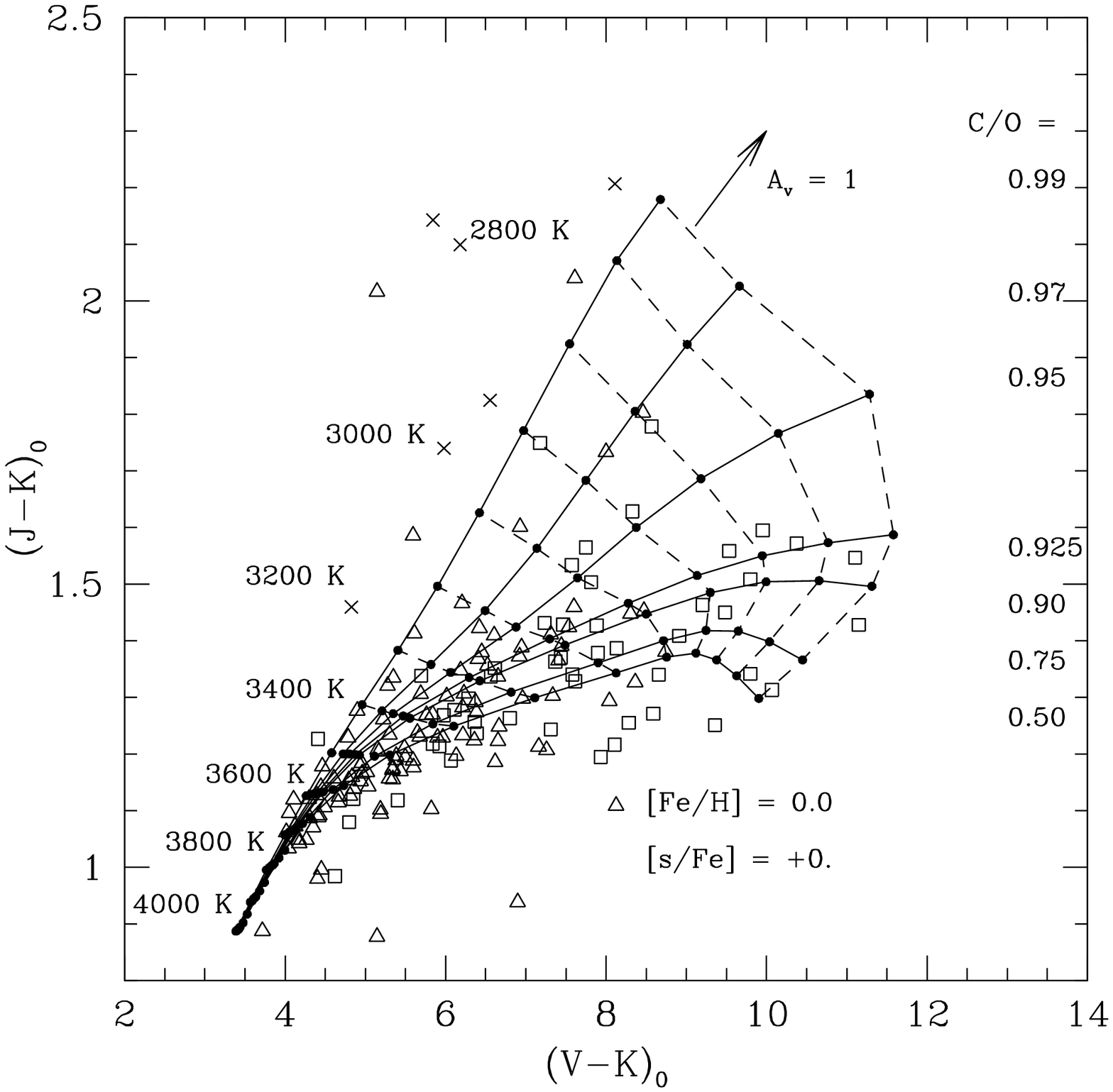}{5cm}{0}{31}{31}{-195}{-50}
\plotfiddle{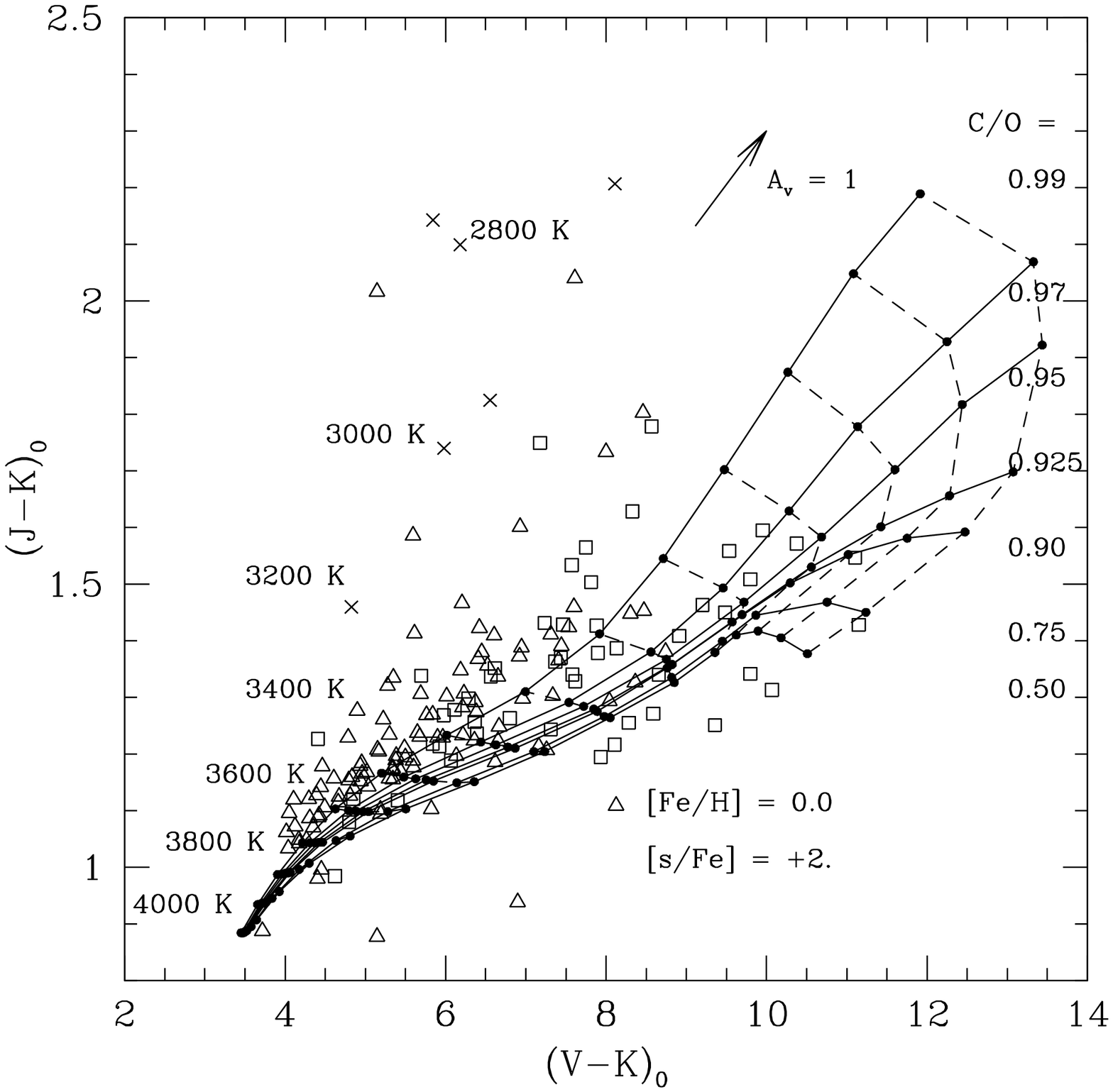}{0cm}{0}{31}{31}{-7}{-24}
\plotfiddle{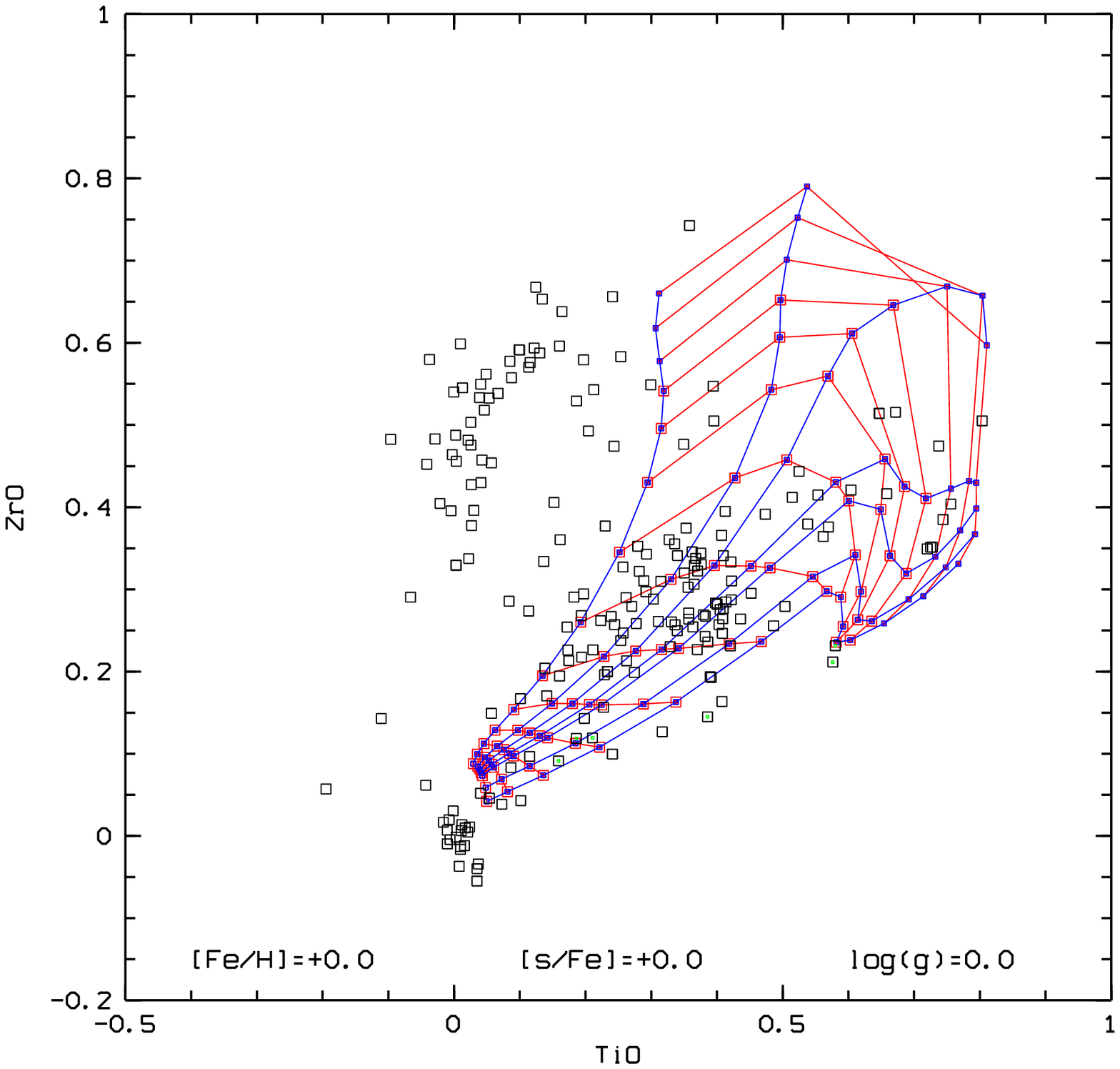}{4cm}{0}{35}{35}{-210}{-140}
\plotfiddle{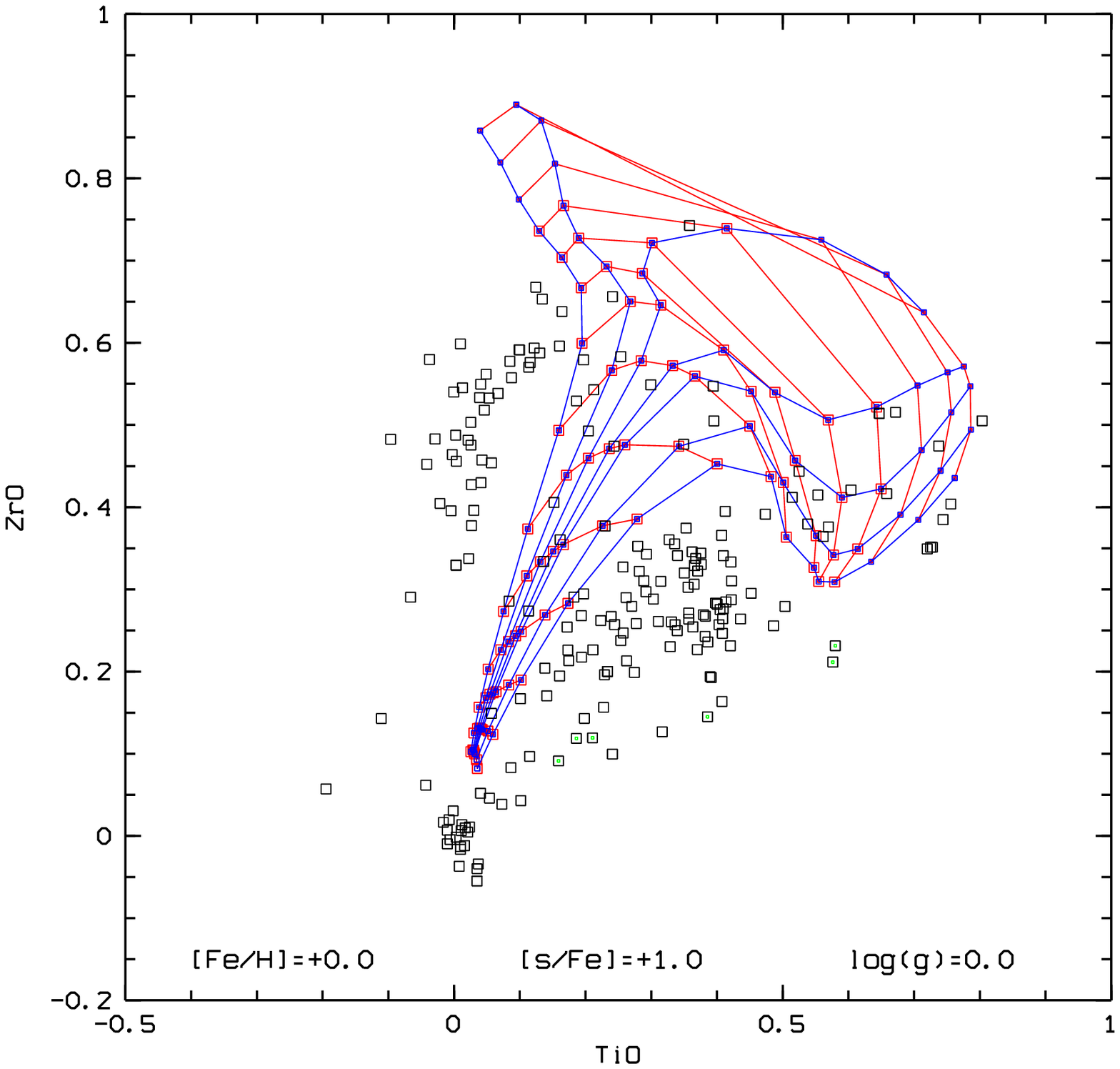}{0cm}{0}{35}{35}{-20}{-116}
\caption{Top panels: Comparison between color indices of observed M (squares), S
  (triangles) and C (crosses) stars, and color indices computed on
  synthetic spectra of S stars for $[$s/Fe$]$ = 0 (left) and 2 (right). The
models with the lowest temperature (2700K) and highest C/O ratio (0.991) are on the top
of each ``grid''. 
Bottom panels: ZrO index versus TiO index for [s/Fe] = 0 (left) and 1 (right). The
  grid corresponds to solar-metallicity, log g=0  models ranging from T$_{\rm eff}$=4000K, C/O=0.5 (around  coordinates 0.05, 0.05 on the leftmost figure) to T$_{\rm eff}$=2700K, C/O=0.99 (around 0.3,0.65). Stars clumping around (TiO,ZrO) = (0,0)
are G and K giants. All S stars to the left of the region covered by the grid are SC stars}
\end{figure}

Synthetic spectra are now compared to observations.
The Henize sample of S stars \citep[205 stars with $ R \le 10.5$ and
$\delta \le  - 25^\circ$;][]{Henize-60} is of particular interest for that
purpose, since (i) it collects S stars with no bias  against high galactic
latitudes \citep{VanEck-Jorissen-99c}, and (ii) a large observational
material has been collected for this sample \citep{VanEck-Jorissen-99b}.
From these data, the $(V - K)_0$, $(J - K)_0$
color-color diagram, dereddened according to \cite{Drimmel-03}, has been
constructed (Fig.~\ref{Fig:TiOZrO}). Similarly, a set of TiO and
ZrO band-strength indices have been computed from the
low-resolution spectra, and displayed on Fig.~\ref{Fig:TiOZrO}.
Their comparison with the model values make it possible
to estimate T$_{\rm eff}$, C/O and [s/Fe] since:
(i) the $(V - K)_0$, $(J - K)_0$ color-color diagram disentangles
T$_{\rm eff}$ and C/O;
(ii) the (TiO, ZrO) diagram disentangles T$_{\rm eff}$ and [s/Fe].
In both cases, there is a good segregation between M and S stars with, however,
some degeneracy between C/O and [s/Fe], especially for low T$_{\rm eff}$.

The $(V - K)_0$, $(J - K)_0$ color-color diagram reveals
that, for a given $V-K$, the range in T$_{\rm eff}$ covered by
models of different C/O ratios can be as large as 400 K. Therefore,
the application to S stars of the usual M-star temperature
scale based on the  $V-K$ index (as done in the past when
specific S-star models were unavailable) leads to errors on
T$_{\rm eff}$ of up to 400 K. 

\section{The atmospheric parameters of S stars}
\begin{figure}[!ht]
\label{Fig:histo-teff-co}
\plotfiddle{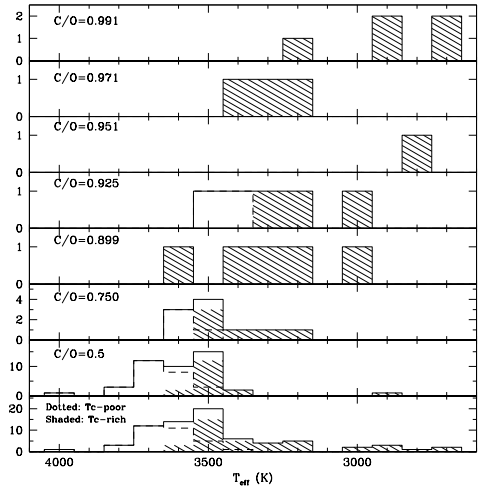}{5.5cm}{0}{35}{35}{-130}{-65}
\caption{Comparison of the T$_{\rm eff}$ distribution of Tc-rich (shaded
histogram) and extrinsic (unshaded histogram). The 7 top panels separate the stars according to the C/O ratio
}
\end{figure}

We have built a ``best model finding tool'', thanks to
an appropriate weighting of well-chosen photometric and narrow-band indices
(dereddened Geneva and VJHKL photometry, ZrO, TiO and NaD band strengths) and chi-square
minimization between observed and synthetic indices.
The adequacy of the selected models has been checked on low-resolution
spectra, dereddened according to \cite{Cardelli-89}; the agreement is very good in most cases (see Neyskens et al., this volume).

Fig.~\ref{Fig:histo-teff-co} presents the distribution of Henize S giants in
terms of temperature and C/O ratio.   The temperature difference between Tc-poor (polluted binary) S stars
  and the cooler Tc-rich (genuine TPAGB) S stars is
  clearly visible. Among Tc-rich stars, despite the small number
  statistics, the expected gradual increase of the C/O ratio as
  the star cools down and ascends the TPAGB is also visible.

This new grid of model atmospheres is an unavoidable prerequisite to reliable spectroscopic chemical analyses
of these objects enriched in s-process nucleosynthesis products. It will
allow us to pursue on a more
quantitative basis the comparison between extrinsic and
intrinsic S stars initiated by \cite{VanEck-Jorissen-00}.

\acknowledgements SVE is FNRS Research Associate and
PN benefits from a fellowship ``Boursier FRIA'' (Belgium).

\bibliography{svaneck}

\end{document}